\documentclass{appolb}
\usepackage{epsfig}

\begin{document}
\title{Shadowing and Fragmentation\\in High-Energy Photo-Mesonproduction%
\thanks{Work supported by DFG and BMBF}%
}

\author{Ulrich Mosel\thanks{mosel@theo.physik.uni-giessen.de}
\address{Institut fuer Theoretische Physik, Universitaet Giessen,
D-35392 Giessen, Germany}
} \maketitle
\begin{abstract}
A model for the description of photonuclear reactions in the
multi-GeV range is described that combines the coherent initial
state interactions (shadowing) of the incoming photon with an
incoherent coupled channel description of final state
interactions of outgoing particles. Particle production itself is
treated either, at low energies, through nucleon resonance decays
or, at higher energies, through string fragmentation. The initial
state nucleons can be far-off-shell due to ground state
correlations. Results are shown for semi-inclusive meson
production.
\end{abstract}

\section{Introduction}
Photo- and Electroproduction of mesons in nuclei offers the
possibility to study the interaction of photons with nucleons in
the nuclear medium. In the energy regime from a few hundred MeV
up to a few GeV one passes the (fuzzy) borderline between a
purely hadronic and a parton-dominated world. One may thus gain
insight into how the transition from meson-production through
'classical' $t$- and $s$-channel processes to the fragmentation
region is achieved; the former processes involve interactions
between nucleons and their resonances, mesons and photons, the
latter is dominated by primary interactions of the incoming photon
with the partons inside the nucleons. How fragmentation is
influenced by surrounding nucleons is so far unknown, but the EMC
effect has shown that inclusive processes of virtual photons with
nucleons change in the nuclear environment \cite{EMC}. From a
study of semi-inclusive processes one may also learn something
about the in-medium properties, in particular their masses, of
the produced hadrons in the nuclear medium. Such information
would be a very valuable supplement to similar studies in
relativistic heavy-ion collisions. While there significantly
higher peak densities are achieved, the reactions proceed through
a long history of densities and temperatures and all observables
integrate intrinsically over this history. In contrast, in
photonuclear reactions the density of the nuclear system is low
($<\rho_0$), but the reactions proceed in a rather stable
environment much closer to the groundstate. In addition, the
in-medium sensitivity of such reactions is as large as that of
ultrarelativistic heavy-ion collions \cite{MoselHirsch}.

Photoproduction of particles on nuclei is usually treated within
the Glauber model (see the review in \cite{Bauer}). While this
model gives an efficient description of coherent initial state
interactions (shadowing), it is rather superficial in its
treatment of final state interactions of the produced hadrons.
The Glauber method -- at least in its practical implementations
-- describes only final state absorption. This may not be
sufficient for a reliable description of semi-inclusive reactions
where the specific hadron-channel under study can not only be
attenuated, but also be fed by interactions with other channels.

In this talk I will present a model that combines the Glauber
treatment of initial state interactions with an incoherent
coupled-channel description of final states. The latter is based
on a coupled channel treatment of hadron interactions in a dense
environment, as it was originally developed for the description
of relativistic and ultrarelativistic heavy-ion collisions. The
model allows to take the important ground state correlations of
nucleons into account and thus to use realistic nuclear
ground-state spectral functions. Particle production in this
model is treated either through a realistic, fixed-by-experiment
description of meson production on nucleons or -- at higher
invariant energies above roughly 2 GeV -- by the LUND string
fragmentation model. The description presented here thus has all
the essential ingredients for a description of high energy
photonuclear processes on nuclei. The details of the model as
applied to photomeson production on nuclei can be found in
\cite{Effehighgam}; it can easily be generalized to
electroproduction (for first results in the lower energy (up to a
few 100 MeV) region see \cite{Lehr}).

\section{Model Ingredients}
Here I briefly summarize the main ingredients of the present
model. Details can be found in \cite{Effehighgam}, \cite{Lehr}
and \cite{Lehrspectral}.

\subsection{Shadowing}
The initial state interactions of the incoming photon with the
nucleus are dominated at the higher photon energies by a coherent
multiple scattering of the photon from the nucleons. The total
photon-nucleus cross section is then given by \cite{Bauer}
  \begin{eqnarray}                   \label{Glauber}
   \sigma_{\gamma A}
   &=& A \, \sigma_{\gamma N}
   + \sum_V \frac{8 \pi^2}{k k_V}
   \Im \left\{ i f_{\gamma V} f_{V \gamma} \int d^2b
   \int_{-\infty}^{\infty} dz_1 \int_{z_1}^\infty dz_2 \,
   n_{12}(\vec{b},z_1,z_2) \right.
   \nonumber \\
   & & \times \left. \e^{i q_V(z_1 - z_2)} \exp\left[ -\frac{1}{2}
   \sigma_V (1 - i \alpha_V) \int_{z_1}^{z_2} dz'\,
   n(\vec{b},z')\right] \right\} ~.
  \end{eqnarray}
Here $\sigma_{\gamma N}$ is the total photon-nucleon cross
section, $f_{\gamma V}$ the amplitude for photoproduction of a
vector meson $V$ on a nucleon, $n$ the nuclear density, $n_{12}$
the correlated two-body density, $q_V$ the momentum transfer from
photon to vector meson, $\sigma_V$ the total vector meson cross
section on a nucleon and $\alpha_V$ the ratio of the real to the
imaginary part of the amplitude for forward scattering of a
vector meson on a nucleon. While it is known since about 25 years
that such a description works well for high energies ($> 5$) GeV,
we have shown that the same expression also describes the
recently observed shadowing at low energies, from about 1 GeV
photon energy on upwards, if the real part of the $VN$ scattering
amplitude is taken into account \cite{Falter}. In \cite{Falter}
we have furthermore shown that this real part leads to an
effective shift of the vector meson mass in the nuclear medium.
Thus shadowing is directly tied in to the ongoing research on
properties of vector mesons in nuclei (for a recent review see
\cite{CassBrat}).

\subsection{Particle Production}
The primary production of a meson $m$ on a nucleon $N$ in the
nuclear medium is in our method described by
  \begin{equation}       \label{mesprod}
   \sigma_{\gamma A \rightarrow A^* m} = \int d^3x \, n(\vec{x})
   s_N(\vec{x}) \sigma_{\gamma N \rightarrow N m}~.
  \end{equation}
Here the \emph{shadowing function $s_N(\vec{x}$)} is implicitly
defined by the total photoabsorption cross section on nuclei
(\ref{Glauber}) written in the form
  \begin{equation}           \label{Glauber1}
   \sigma_{\gamma A} = \int d^3x \, n(\vec{x})
   s_N(\vec{x}) \sigma_{\gamma N}
  \end{equation}
and $\sigma_{\gamma N \rightarrow N m}$ is the free cross section
for producing meson $m$ on the nucleon $N$. Equation
(\ref{mesprod}) essentially corresponds to multiplying the total
cross section in (\ref{Glauber1}) with the partial cross section
for producing meson $m$.

For invariant energies in the photon-nucleon system below 2.1 GeV
we obtain the needed cross sections for the primary interactions
$\gamma N$ as in \cite{EffeHom} by an explicit calculation of the
cross sections for producing nucleon resonances as well as
one-pion, two-pion, eta, vector meson and strangeness. For higher
energies where specific nucleon resonances are no longer visible
we use the string fragmentation model FRITIOF \cite{Lund} in the
way described in \cite{Effehighgam}.

\subsection{Final State Interactions}
The final state interactions are described in a semiclassical
coupled channel transport calculation that includes all nucleon
resonances up to an invariant mass of 2.1 GeV, all relevant
mesons in this range, and the string fragmentation mechanism of
the FRITIOF model for higher energies, consistent with the
description of the primary production process. The model,
described in detail in \cite{Effegam}, allows a full coupled
channel treatment of final state interactions, including
nonforward processes. It has been tested and proven to give a
very good description of photonuclear processes in the MAMI
energy regime (up to 800 MeV photon energy) \cite{Lehr,EffeHom}.

\subsubsection{Groundstate Correlations}

\begin{figure}
\begin{center}
\epsfig{file=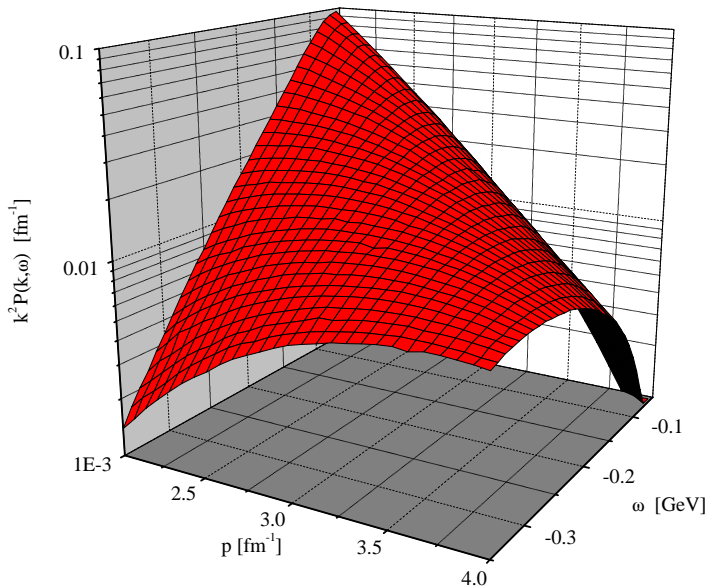,width=12cm}
\end{center}
\caption{Energy and momentum dependence of the nucleon spectral
function P(k,$\omega$) for energies below the Fermi energy
$\omega_F$ (from \protect\cite{Lehrspectral}).} \label{specfunc}
\end{figure}

Nuclear groundstate correlations manifest themselves in the
presence of high-momentum components, far beyond the Fermi
momentum of nuclear matter, in the nuclear wavefunctions. Their
presence can have significant consequences for particle
production close to or even below the free threshold. While there
are very sophisticated many body calculations (see e.g.\
\cite{Benhar}) of these correlations we have recently shown (in
\cite{Lehrspectral}) that the spectral function of nucleons in
nuclear matter is dominated by phase-space effects and can be
calculated from exactly the collision integrals entering the
transport calculations just described. An example of these
results is shown in Fig.\ref{specfunc}.

We have recently implemented this spectral function into our
transport calculations. These spectral functions have a strong
influence on the primary particle production very close to
threshold \cite{Sibirtsev}. In the GeV range treated here, this
would be the case, for example, for $D$-meson or $J/\psi$
photoproduction. In addition, the spectral functions may
influence the FSI in the first few reaction steps.

\section{Results}
\begin{figure}
\begin{center}
\epsfig{file=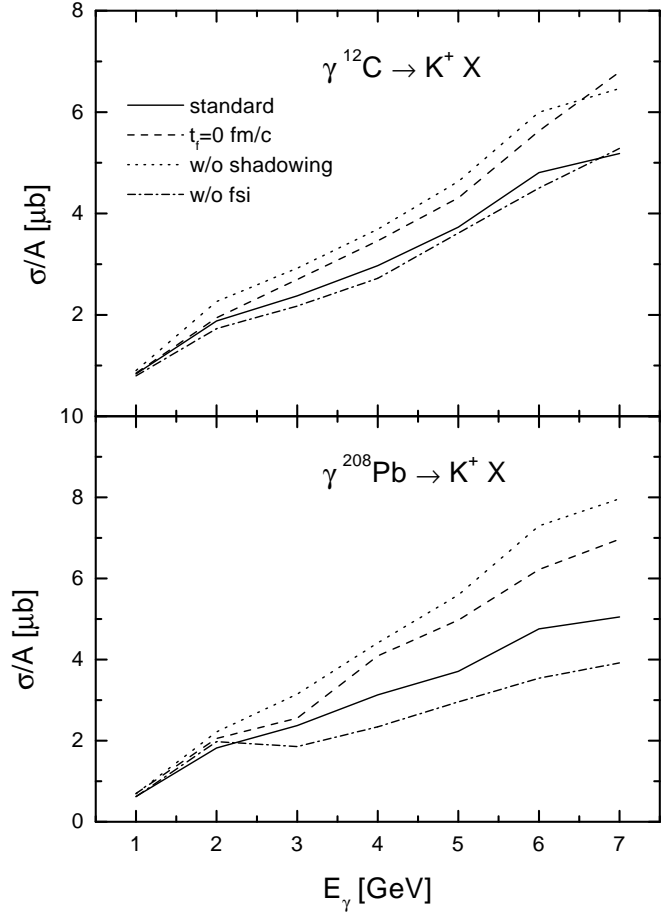,width=10cm}
\end{center}
\caption{Total cross section for $K^+$ production in $\gamma C$
(top) and $\gamma Pb$ (bottom). The dotted curve shows the result
without shadowing, the dashed curve illustrates the effect of
setting the formation time in the string fragmentation model to
zero, the solid curve gives the results of the full calculation
and the dash-dotted curve those without final state interactions}
\label{K+exc}
\end{figure}

In \cite{Effehighgam} I show results of this approach for pion,
eta, kaon and $\pi^+\pi^-$ production. Here I just demonstrate the
effects of various model ingredients on kaon photoproduction. In
Fig.\ref{K+exc} the total cross sections for photoproduction of
$K^+$ mesons are shown. These results show that shadowing is much
more effective in the heavier nucleus $Pb$ than in $C$, becoming
important for energies $> 2$ GeV. At the highest energy shown
here (7 GeV) it reduces the cross section to about 60\% of its
unshadowed value. The formation time of the string fragmentation
model has a somewhat larger effect in the heavier nucleus because
there reabsorption has more time to act in the case of zero
formation time (the standard value used is 0.8 fm/c). The most
interesting effect is the \emph{enhancement} of the $K^+$
production cross section \emph{due to final state interactions}.
This is a clear coupled channel effect: since the $\bar{s}$ quark
in the $K^+$ meson cannot be absorbed in collisions with nucleons,
there is only very little reabsorption of low-energy $K^+$ mesons.
The final state interactions then act only to populate this
channel through secondary interactions, such as, e.g., $ \pi^+ N
\rightarrow K^+ \Sigma$.
We expect a similar behavior for the charmed $\bar{D^0}$ and
$D^-$ mesons with a free threshold of about 3.7 GeV.

Thus, the full coupled channel treatment of final state
interactions is important for all those particles whose mean free
path in matter is long. Here the sidefeeding has to be taken into
account.

\begin{figure}
\begin{center}
\epsfig{file=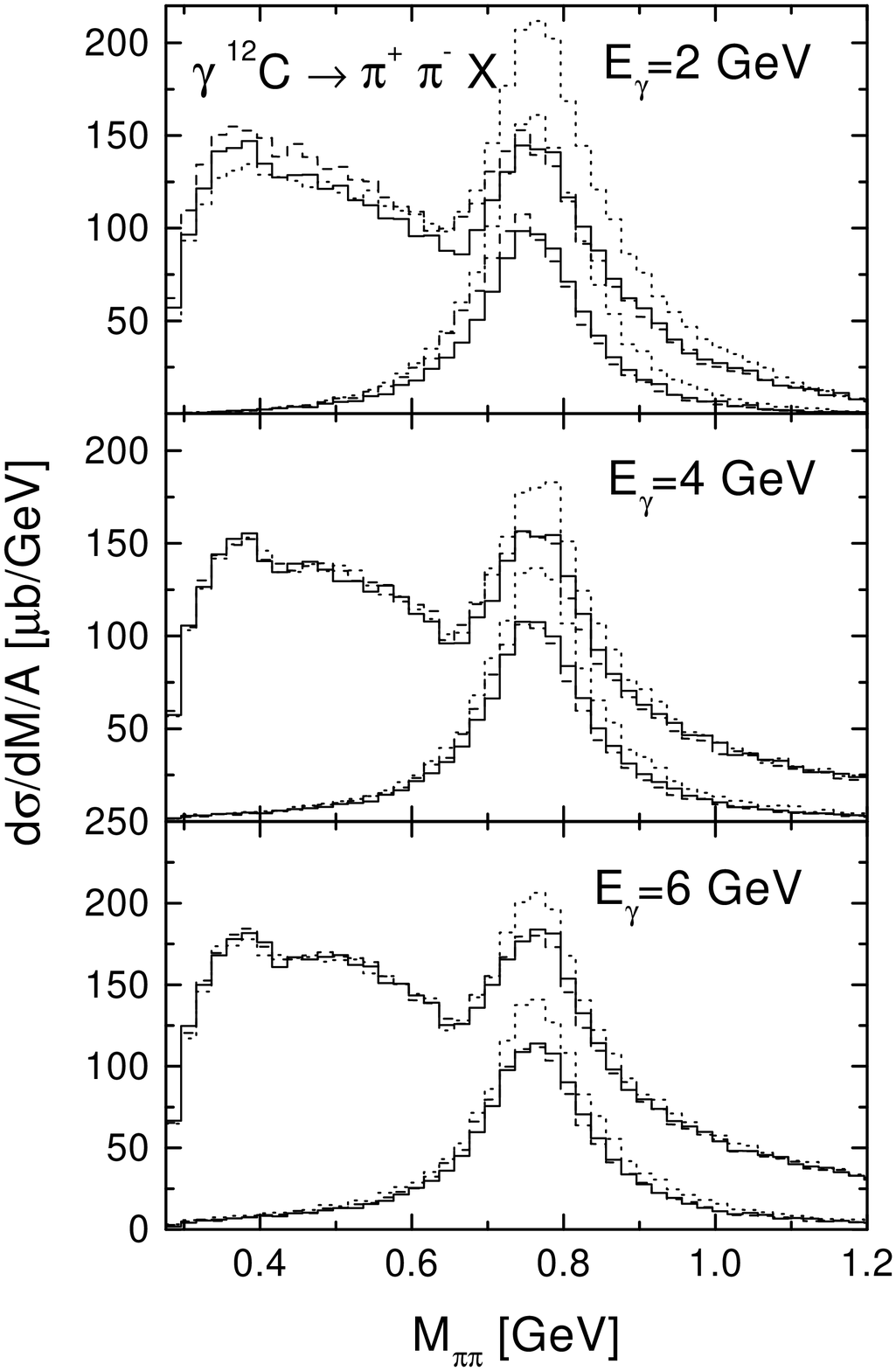,width=10cm}
\end{center}
\caption{Invariant mass spectrum of inclusive $\pi^+\pi^-$
production in $\gamma C$ reactions. The dashed lines give the
results of a dropping mass scenario for the $\rho$ meson. The
dotted lines are obtained without any final state interactions.
The total mass-differential cross sections as well as the
contributions of the $\rho$ are shown.} \label{rho}
\end{figure}

As a second example of the results obtained with this method I
show in Fig.\ \ref{rho} the invariant mass distribution of
$\pi^+\pi^-$ pairs for bombarding energies of 2 - 6 GeV on a $C$
target. This reaction is of interest in connection with the
$\rho$-electroproduction in high-energy reactions \cite{Hermes}.
In the calculations presented here we know which pions originated
from a $\rho$; we can thus distinguish between true and random
coincidences.  The final state interactions (FSI) play a dominant
role at the lowest energies whereas their effect becomes smaller
at the highest one. This is due to the energy-dependence of the
$\pi N$ cross section. It is also seen that the $\rho$ meson can
be identified in the mass-spectrum at all energies; this becomes
more difficult in heavier nuclei. The strong background rising
down to the threshold at $2 m_\pi$ is due to pions from resonance
decays. The figure also shows that a mass shift of the $\rho$
meson according to the prediction of \cite{Hatsuda}
  \begin{equation}
   m^*_\rho = \left( 1 - 0.18 \frac{\rho}{\rho_0} \right) m_\rho
  \end{equation}
is hardly visible in the pion invariant mass spectrum. This is
simply due to the strong reabsorption and rescattering of pions
in the nuclear medium; the $\rho$ peak finally seen is then that
of a $\rho$ meson already decaying in the nuclear surface.

\section{Summary}
In this talk I have described a new theoretical approach to
photo- and electroproduction at high energies. This approach
combines shadowing in the incoming state with a coupled-channel
treatment of final state interactions. The FSI are treated in a
transporttheoretical method, originally developed for the
description of heavy-ion reactions, and later also extended to
hadron-nucleus \cite{Effepi} and photo- and electro-production
experiments at smaller energies \cite{Lehr,EffeHom,Effegamma}
where it yielded very good results in comparison to MAMI data. I
have illustrated some results of this method for photoproduction
in the GeV energy regime; more can be found in a recent
publication \cite{Effehighgam}.

The method, when extended to higher energies and
electroproduction, shows some promise for investigations of
phenomena such as color transparency and the EMC effect. Since
the model also allows to include realistic nuclear spectral
functions in a simple way \cite{Lehrspectral} we expect it also
to be useful for a quantitative, realistic description of
semi-inclusive reactions on nuclei. It should also help in a
reliable analysis of experiments aiming for a determination of
nuclear structure functions and their transition to partonic
degrees of freedom, since the model contains the transition from
nucleon-resonance dominated particle production at lower energies
to a string fragmentation production at the higher energies.

\section{Acknowledgments}
I gratefully acknowledge many helpful discussions with M.
Effenberger, on whose thesis a large part of this talk is based,
and with J. Lehr, H. Lenske and S. Leupold. This work was
supported by DFG and BMBF.


\begin{thebibliography}{99}

\bibitem{EMC} D.F. Geesaman, K. Saito, and A.W. Thomas, Ann. Rev.
Nucl. Part. Sci. \textbf{45}, 337 (1995)

\bibitem{MoselHirsch} U. Mosel, in: Proc. Baryons'98, World
Scientific, Singapore 1999, p. 629

\bibitem{Bauer} T.H. Bauer et al., Rev. Mod. Phys. \textbf{50}, 260 (1978)

\bibitem{Effehighgam} M. Effenberger and U. Mosel, Phys. Rev. \textbf{C62},
014605 (2000)

\bibitem{Lehr} J. Lehr, M. Effenberger, and U. Mosel, Nucl.Phys. \textbf{A} 671, 503
(2000)

\bibitem{Falter} T. Falter, S. Leupold, and U. Mosel, nucl-th/0002062

\bibitem{CassBrat} W. Cassing and E.L. Bratkovskaya, Phys. Rep.
\textbf{308}, 65 (1999)

\bibitem{EffeHom} M. Effenberger et al., Nucl. Phys. \textbf{A614}, 501 (1999)

\bibitem{Lund} B. Anderson, G. Gustafson, and Hong Pi, Z. Phys.
\textbf{C57}, 485 (1993)

\bibitem{Effegam} M. Effenberger, E.L. Bratkovskaya, U. Mosel,
Phys. Rev. \textbf{C60}, 044614 (1999)

\bibitem{Benhar} O. Benhar, A. Frabrocini and S. Fantoni, Nucl.
Phys. \textbf{A505}, 267 (1989); Nucl. Phys. \textbf{A550}, 201
(1992)

\bibitem{Lehrspectral} J. Lehr et al., Phys. Lett. \textbf{B483}, 324 (2000)

\bibitem{Sibirtsev} A. Sibirtsev , W. Cassing, and U. Mosel, nucl-th/9909028

\bibitem{Hermes} Ackerstaff et al., Phys. Rev. Lett \textbf{82},
3025 (1999)

\bibitem{Hatsuda} T. Hatsuda and S. Lee, Phys. Rev. \textbf{C46},
R34 (1992)

\bibitem{Effepi} M. Effenberger et al., Phys. Rev. \textbf{C60},027601
(1999)

\bibitem{Effegamma} M. Effenberger et al., Nucl. Phys. \textbf{A613}, 353
(1997)

\end{thebibliography}
\end{document}